# Ethics in Computing Education:
# Challenges and Experience with Embedded Ethics

Sudeep Pasricha
Department of Electrical and Computer Engineering, Colorado State University
Fort Collins, CO, USA
sudeep@colostate.edu

## ABSTRACT
The next generation of computer engineers and scientists must be proficient in not just the technical knowledge required to analyze, optimize, and create emerging microelectronics systems, but also with the skills required to make ethical decisions during design. Teaching computer ethics in computing curricula is therefore becoming an important requirement with significant ramifications for our increasingly connected and computing-reliant society. In this paper, we reflect on the many challenges and questions with effectively integrating ethics into modern computing curricula. We describe a case study of integrating ethics modules into the computer engineering curricula at Colorado State University.

## 1. INTRODUCTION

Computing technologies are today deeply intertwined with our everyday lives. Smart embedded and IoT computing devices are at the heart of digital platforms that we use to interact with other individuals, commute, work, prepare food, and enjoy leisure time. More broadly, these computing devices are being increasingly used in the industrial automation, healthcare, telecommunication, and avionics sectors. While the rapid proliferation of computing technology has improved our lives in many ways, it has also led to harmful consequences [1]. Alarming privacy violations have been reported from smart TVs and smart speakers that spy on users without consent [2] and toys with cameras that spy on vulnerable children [3]. Increasing rates of recalls of semi-autonomous vehicles [4] and medical devices [5] due to security and safety vulnerabilities are starting to reduce confidence in these technologies. Artificial intelligence (AI) and machine learning algorithms that demonstrate bias during criminal justice decisions [6], allocation of scarce healthcare resources [7], and employment decisions [8] are receiving increased scrutiny from the public, media, and regulatory agencies. Reports of semiconductor fabrication of computing chips involving the use of toxins that cause health issues amongst workers [9] and the use of forced labor [10] have raised widespread concerns. In the wake of these scandals, the need to train computing professionals who are aware of the consequences of the actions that they take and are equipped to make ethical decisions is more urgent than ever before.

But first, *what exactly does it mean to be an ethical computing professional?* One can think of such an individual as conducting technology design that avoids bias and other harmful practices, with conscious knowledge of the goals and politics surrounding the technology, and an intent to ensure that the consequences of deploying the technology result in good and fair outcomes. What constitutes "good" is of course not without debate. Even the term "ethics" often invokes confusion, and it is difficult to disassociate it with the related broad concepts of values, justice, and responsibility. Further complicating matters is the widely held notion that being ethical is synonymous with complying with the law, which is increasingly not true, as the rate of technological advancements generally outpaces the ability of law to keep up [11]. Thus, computing professionals today face a lot of uncertainty when deciding between competing demands of maximizing efficiency and profit, supporting accessibility for diverse populations, and avoiding harm. The complex ethical dilemmas associated with the myriad design decisions across hardware and software design processes are anything but easy to resolve.

To address the growing public dissatisfaction over unethical developments in computing and technology, particularly surrounding ubiquitous social media and AI deployments, the computing industry is responding by hiring experts in ethics to provide oversight and outline best practices during product development and deployment [12]. But such top-down efforts are often disconnected with analyses and implementations being performed by early-career computing professionals making low-level design decisions, who often are unaware of the ethical implications of technologies that they work on or develop. Ethics education can provide students with the foundational skills and the motivation to make ethically informed decisions and to incorporate ethical considerations in the design and deployment of computing technology. The soon to be early-career computing professionals can learn about not only what computing technology they can create and how to solve technical challenges related to it, but also whether they should create that technology, and how best to realize their technology implementations in the least harmful way. However, the ability to think in such a way requires careful interventions in computing curricula at universities.

In this paper, we explore the current state of the art with ethics education in the computing field. We outline challenges with teaching and integrating ethics content in contemporary computer engineering and computer science curricula. Then, we discuss our pedagogical experiences with integrating ethics modules into the computer engineering curricula at Colorado State University. We also reflect on content and student feedback, with a goal of sustainably integrating ethics into modern computing curricula.

## 2. BACKGROUND

Even though ethics has been a topic of much debate and fascination over the millenia, one of the earliest emphases on ethics with respect to engineering and sciences was in the 1965 book titled "Ethical Problems in Engineering" [13]. Formal courses on ethics in engineering and computer science were not taught until the late 1970s and early 1980s. This was the time when professional organizations in the field realized the need for developing ethical competencies among computing and engineering graduates and laid the foundation for computing ethics education [14].

Starting from around 2000, ABET, the organization that accredits undergraduate engineering programs at universities in the United States, has required students to be trained to have an understanding of professional and ethical responsibility. Interest in ethical issues in engineering and computer science has grown significantly since then. Curricular guidelines laid out by ACM and the IEEE Computer Society over the past two decades (Computing Curricula 2001 [15], 2013 [16], and 2020 [17]) have continued to emphasize the importance of computer ethics by requiring core hours of instruction in this area. Many undergraduate engineering and computer science programs now offer modules on ethics in engineering and ethics in computer science as part of their curricula. There are also several active journals on ethics, and even ethics and education, and dozens of textbooks on the topic. Many conferences and workshops have dedicated sessions on ethics, which today are often dominated by discussions of ethical concerns with the emerging areas of AI and autonomous systems.

Current computer science and engineering programs typically cover the topic of ethics by teaching standalone computer ethics courses within their department [18] or requiring students to take an ethics course taught outside the department, often in Philosophy [19]. Researchers working on computing education have raised concerns about such traditional approaches to teaching computing ethics. They argue that offering stand-alone computing ethics courses gives students the impression that ethics is not truly an integral part of the computing domain, but rather a separate and specialized task that is someone else's responsibility [20].

Regarding the question of how best to teach computing ethics, several strategies for ethics pedagogy have been outlined within the computing education research community, e.g., gamification [21], immersive theater [22], and incorporation of science fiction [23]. However, the manner in which different universities and programs teach ethics remains far from homogeneous in both extensiveness and content [24]. Studies have also shown that ethics instruction often does not translate into experiences outside the classroom [25]. Therefore, there are still many unanswered questions about how to best accomplish the goals of ethics education, particularly in the computing field.

## 3. COMPUTING ETHICS: KEY QUESTIONS

In this section, we discuss what we believe are some of the most pressing questions that are relevant for the effective teaching of ethics in university computing curricula.

**Can ethics be taught?**

Despite the many efforts towards integrating ethics into curricula across computer science and engineering departments in universities, perhaps we should start with a very fundamental question: *is it even possible to teach computing ethics?* There has been debate on whether teaching ethics in college is a futile effort. Karl Stephan, an engineering professor, described the following encounter he once had with his colleague [26]: "Some years ago I argued with a fellow professor about the issue of engineering ethics education at the college level. His point was along the lines of, 'Hell, if eighteen-year-old kids don't know right from wrong by the time we get 'em, they're not going to learn it from us.'

At the heart of the question is the manner in which we make moral decisions and whether that is something innate and fixed, or something that can be influenced. Social intuitionists claim that humans rely to a large extent on their intuitions when making moral decisions. They say that moral reasoning comes in only after intuitions, to explain and potentially justify what intuitions tell an individual to do. Jonathan Haidt summarized this as follows, "The central claim of the social intuitionist model is that moral judgment is caused by quick moral intuitions and is followed (when needed) by slow, ex post facto moral reasoning" [27]. Social intuitionists give the example of how some people are repulsed by the ideas of incest or torture, or eating animals, and thus their ethics is based on their intuition and not reasoning. In other words, they claim that ethical behavior is something primitive within humans that is not amenable to reason, and thus change.

However, there are serious criticisms of the social intuitionist viewpoint [28]. One argument is that moral intuitions do not come out of nowhere, but in fact, people acquire intuitions through developmental processes that involve an individual's experiences and reasoning about those experiences. If this is accurate, then ethics education can lead to better moral intuitions. Another argument against social intuitionism is that people can be convinced via discussion and debate with their friends to change their intuitions, e.g., be convinced to become vegetarian or refrain from hate speech that vilifies certain races and genders. Thus, it is possible to influence ethical decision making with reasoning. Yet another argument points to more complex situations, where intuitions may not be very helpful. For instance, consider the case when an engineer must decide whether to alert authorities about the potentially illegal behavior of her manager, or to listen to a colleague, who is a good friend, and does not want her to blow the whistle on the manager. In such a case, the engineer must think carefully about the consequences of her actions. Reasoning about the actions to take, in this situation, is not just helpful but essential.

If we agree with these arguments, we can also think about how a computing professional may require information and knowledge to help them make ethical decisions. This is similar to how knowledge is required to make technical decisions in computing. Imparting this technical knowledge is in fact the focus of most computing curricula today. Similarly, ethical knowledge can also be imparted, to help computing professionals understand professional norms and practices, learn to identify ethical issues and make ethical decisions, and comprehend the social consequences of technical endeavors. Ethics education can thus increase the likelihood that computing professionals will be better able to handle ethical issues that arise in their careers.

**How do ethical theories fit into computing ethics curricula?**

Humans have debated how they should treat one another, what constitutes right and wrong, and good and bad, for thousands of years. These discussions are at the heart of ethics and also why it is a normative domain as it does not focus on what does happen but rather on what should happen. For such normative matters, we can learn from philosophers (Western and Eastern) who have systematically explored questions of morality in great depth. Three ethical theories have been the focus of most attention: utilitarian ethics, Kantian ethics, and virtue ethics. Each theory demonstrates a way in which ethical analysis can lead to better understanding of complex situations that may arise in our everyday lives.

Utilitarianism [29] is a philosophy in which a behavior is considered right or wrong based on its consequences. Thus, it can be considered a consequentialist theory. According to the classical utilitarianist approach, actions are considered good when they produce the most happiness and bad when they produce net unhappiness. Therefore, a person deciding what to do should weigh different options according to their consequences and pick the action that will bring about the greatest amount of positive consequences (e.g., most overall net happiness in the world). It is possible that the right action may be the one that brings about some negative outcomes (such as unhappiness), but that action is still considered justified if it leads to the most net happiness or the least net unhappiness out of all the alternatives. It is also possible that the right action will diminish the happiness of the person taking the action because that is what will increase overall happiness. The theory of utilitarianism has been widely influential among legislators and public policy makers who seek out policies that will have a net positive impact, despite some negative consequences. Computing professionals can use this theory for cost-benefit and risk-benefit analysis to understand positive and negative effects of actions on clients, employers, and the public, before calculating which action would be better. But even though utilitarianism provides a reasoned and systematic framework to think about ethics and morality, it has shortcomings. Critics argue that the theory tolerates, and even recommends, that burdens be imposed on some individuals for the sake of overall good. Thus enslaving a few individuals to benefit a larger number of individuals is justified under the utilitarianist theory.

According to Kantianism [30], actions that are done from a sense of duty are morally worthy, and those done for self-interest or simply out of habit are not. As this theory emphasizes acting from duty, it is considered a deontological theory. According to the theory, morality can be considered as a set of rules that humans adopt to direct their behavior. One can think of certain fundamental ethical rules that must be followed by all humans, e.g., respecting people's privacy and respecting people's right to make decisions about their own lives. This implies that certain actions, e.g., killing people or slavery, are clearly prohibited. Thus, in Kantianism, the consequences of actions are not as important as what a person intends to achieve by an action. According to the theory, one should never treat another individual merely as a means, rather one should always treat others as ends in themselves. This is because by treating someone merely as a means to one's own ends, we deny their humanity and also deny their capacity to make choices and reason for themselves. This is what is particularly problematic with slavery and other utilitarian practices that treat some human beings as a means to increase the happiness of other human beings. Computing professionals can learn from the Kantian emphasis on universal human rights, and the obligation to treat all stakeholders with respect. A desire to increase profit should not come at the cost of these obligations.

Whereas utilitarianism and Kantianism focus on actions, the theory of virtue ethics [31] is concerned with traits of character. According to virtue theory, ethics has to do with humans, so our attention should be focused on what it means to be a good human being, rather than merely actions. A good person must have some traits to lead a good life, that constitute excellence in a human being. These traits (virtues) could include honesty, courage, and generosity. It is also understood that these virtues are not inborn, but rather must be cultivated through practice, habit, and practical reasoning. Computing professionals can apply the principles of virtue ethics in whatever role they find themselves. When they are in a situation requiring moral judgment or decision-making, rather than simply looking for a rule or decision procedure to follow, they should consider the virtues associated with being a good X, where X is a particular role such as architect, programmer, or manager. Such a way of thinking can direct the professional to an appropriate form of behavior, typically one that considers the safety and welfare of society, while leveraging the virtues of honesty and courage to achieve desirable goals.

Ultimately, computing ethics is concerned with how computing professionals deal with stakeholder relationships (e.g., clients, employers) and considerations of the social impact of their work. For dealing with such ethical aspects, one can draw from one or more of the ethical theories discussed above. These ethical theories provide a way of thinking about ethics that is systematic and disciplined, and therefore they can be very useful to navigate tricky situations in which computing professionals find themselves. The theories provide a structured framework, define key concepts, and provide the appropriate vocabulary to allow professionals to interpret their specific situations, identify ethical dilemmas, and begin to formulate strategies to address them. Without knowledge of ethical theories, professionals may not be able to comprehend complex ethical scenarios fully and deeply. As such, it is important that computing ethics curricula integrates discussions and applications of the rich theory behind ethics.

**Do engineering and computing codes of ethics matter?**

Both the IEEE and ACM have codes of ethics that are relevant for computing professionals. However, these codes have been a subject of much criticism. Many experts question the effectiveness of these codes in computing ethics education. In [32], the authors described how students in their studies found ethical theories to be much more insightful for addressing ethical dilemmas than the professional code of ethics. In [33], it was shown that when study participants that consisted of software engineering students and professional software developers were asked to refer to the ACM code of ethics to respond to ethical dilemmas, it had no effect on ethical decision making when compared to a control group.

One major critique of these codes is that they are too generic and provide no practical guidance on how to act when faced with ethical dilemmas. Another criticism is that these codes are not effective as they lack enforcement power. In certain engineering fields, one must possess a license to provide certain services. This licensing is the responsibility of state licensing boards, and they generally require adherence to a code of ethics. The state board enforces these codes and violators are subject to disciplinary action. An engineer's license can be revoked or suspended, and engineers can even be required to pay penalties if they violate the codes. Unfortunately, in the computing profession, professionals are rarely, if ever, expelled from ACM or IEEE for failure to live up to a code. Moreover, even if a member was expelled from ACM or IEEE, it is unclear if that would impact their career opportunities, as membership to these organizations is not required for most jobs.

We believe that many of the critiques of codes of ethics are somewhat harsh and misguided. When computer engineering or computer science students complete their undergraduate education, they are not just college graduates, but are also engineers and scientists. In these professions, codes of ethics proclaim the commitment of practicing individuals towards using their expert knowledge in ways that serve the public or, at the very least, in ways that do not harm the public. The codes are thus critical for establishing public trust in the computing profession, and setting apart computing professionals as a distinctive group different from others without degrees or with degrees in other fields of higher education. The public trust in turn is crucial towards allowing computer engineers and scientists the degree of autonomy that is needed to effectively apply their expert knowledge to solutions that impact society in innumerable ways.

Beyond their impact on the public trust of the computing profession and individual practitioners, codes of ethics shape the expectations of all those who interact with computer engineers and scientists, including employers, media, government officials clients, contractors, and others. They are thus an integral part of the image and reputation of the computing profession. For individual computing professionals, these codes represent the collective wisdom of members of the profession, painstakingly distilled into a relatively small set of statements describing the principles that computing professionals should embrace. Such guiding statements are particularly valuable for early career computer engineers and scientists in describing what is expected of them, whether they join a startup with unprofessional and sloppy standards, or a more established company where principles of ethics may be held in high regard until they infringe on the company's bottom line, in which case they are treated as an inconvenience. Ultimately, even though codes of ethics do not provide specific guidance on how to resolve specific ethical dilemmas, ethics education should expose computing students to these codes, to help establish early expectations of the computing profession and standards of professional behavior. Even more beneficial is to teach students how to interpret and debate the meaning of the codes, and not to treat them as rules to be followed blindly. This can prevent legalistic thinking, which is problematic as it may suggest that a person should simply follow the rules without carefully considering nuances and dimensions of a situation that are not covered in the wordings of the rules. The ability to examine the implications of codes and practice applying them across diverse scenarios can help students immensely when it comes time to handle ethical dilemmas that involve competing interests and norms in their professional careers.

**What topics should computing ethics education focus on?**

The topics covered in computing ethics modules must have relevance to contemporary ethical challenges in industry, public policy, and research. In a broad qualitative analysis of 115 technology ethics courses across universities, [24] identified high level topics that were covered across the syllabi. The most common topics that occurred in at least a quarter of the courses and their frequency across courses are shown in Table 1.

**Table 1: Number of technology ethics courses across universities that had content for each topic listed, out of 115 total courses [24]**

| Topic | # Courses |
|---|---|
| Law and policy | 66 |
| Privacy and surveillance | 61 |
| Philosophy | 61 |
| Inequality, justice, and human rights | 59 |
| AI and algorithms | 55 |
| Social and environmental impact | 50 |
| Civic responsibility and misinformation | 32 |

The most common topic was *law and privacy*, with an emphasis on comprehending and interpreting law as part of the course objective. Courses that included this content covered policies that specifically impact computing and technology, such as DMCA (Digital Millennium Copyright Act) and GPDR (General Data Protection Regulation), as well as other legal concepts such as intellectual property and free speech, as they apply to computing. The next most common topic was *privacy and surveillance*, where the focus ranged from anonymization within big data and across social media to surveillance in public spaces and with smart mobile devices, with coverage of issues such as our digital footprints and ownership of data. Ethical theories such as utilitarianism, deontology, and virtue ethics were covered under *philosophy* in 53% of the syllabi analyzed. Courses covering this topic discussed cultural norms, the nature of morality, and how to develop a personal code of ethics. Just over half the syllabi covered themes of *inequality, justice, and human rights*, highlighting how technology can amplify societal discrimination, promote marginalization of some populations, and spread inequality. The next most common topic was *artificial intelligence* (AI), with a focus on algorithmic fairness and bias, as well as issues of regulation, explainability with automated decisions, and morality in the context of self-driving cars, autonomous drones, and smart weapons. Several syllabi focused on the large-scale social or environment impact of technology. Courses covering this topic discuss the paradox of technology increasing social connections while also weakening social structures that rely on in-person interactions, challenges with online harassment, doxing, hate speech, and trolling, and the impact of technology on environmental degradation and sustainability. The topic of *civic responsibility and misinformation* appeared in 28% of syllabi. Here the focus was on combatting disinformation, as well as causes of polarization, such as fake news and social media bubbles. Other relatively less frequently occurring topics emphasized ethics during research, codes of professional ethics, cybersecurity, and the future of work and labor.

**Who teaches computing ethics?**

Another important concern has to do with who should teach computing ethics courses. It has been argued that philosophers and social scientists trained in ethics should be teaching these courses, as they have the expertise needed to shepherd debates and lead more informed discussions [19]. However, there have also been strong counterarguments that emphasize how it is critical for computing faculty to teach these courses, so that students understand that ethical issues are a fundamental and integral part of computing, and not a tangential topic that they can take a course on elsewhere. We subscribe to the latter approach. In some cases, there have been efforts to involve both philosophy and computing instructors to teach ethics courses [12]. However, this is a very resource-intensive solution that is difficult to orchestrate due to the competing demands and curricula teaching needs of schools to which each of the instructors belongs.

## 4. CASE STUDY: COMPUTING ETHICS @ CSU

Realizing the importance of ethics in computing, at Colorado State University (CSU), we have launched an initiative to teach ethics as part of short modules across multiple courses in the computer engineering curricula. This approach borrows from the fields of law, medicine, and business that have all considered the benefits of infusing ethics throughout a curriculum instead of covering ethics content in standalone classes [34], [35]. Recently, some other universities have also started similar curricula-wide ethics integration, e.g., Harvard's Embedded EthiCS program [12]. Our approach is somewhat different from existing efforts in the following ways: 1) most recent efforts target integrating ethics into undergraduate curricula, whereas we focus on integrating ethics modules across courses that encompass both undergraduate and graduate courses; and 2) most existing efforts focus on integrating ethics into computer science curricula, whereas we target ethics for electrical and computer engineering, which requires going beyond software ethics and also emphasizing ethical hardware and microelectronics system design.

The seed for our approach was planted many years ago as part of the NSF-sponsored RED (Revolutionizing Engineering and Computer Science Departments) project at CSU that spanned 2015-2020 [36]. The main goal of RED was to provide a holistic education, with knowledge integration across courses with three key threads that extended throughout the curriculum, namely: foundations, creativity, and professionalism. The professional formation thread was designed to impart professional skills that go beyond the traditional technical curriculum for the development of well-rounded engineers, so that they are prepared to enter the workplace. A critical component of this thread involved exposing students to ethical considerations that they may encounter in their professional careers and preparing the students to deal with them.

Unfortunately, the focus of the RED project was on undergraduate electrical engineering students, which not only left out undergraduate computer engineering students (and graduate students in both electrical and computer engineering), but also failed to integrate computing-centric ethics modules for electrical engineering students. Moreover, the effort with teaching ethics to electrical engineering students lasted a very short time, after which those modules were abandoned. Students complained about the modules being too ambiguous and philosophical, and instructors soon reverted to focusing on technical modules that were better appreciated, instead of the ethics modules. The responsibility of teaching students ethics was delegated to mandatory standalone workshops that were presented to our students through the Professional Learning Institute (PLI) at CSU.

Our recent emphasis on ethics integration is meant to be a reboot, where we learn from part mistakes to create a more immersive and engaging ethics integration into the computing curricula. The motivation for teaching ethics arose from our observations during online and in-class discussions in computing courses where students repeatedly failed to demonstrate the ability to reason about and understand ethical issues. For students engaged in technical assignments and labs in these courses, technical concerns seemed to be paramount, and unethical behavior (including various forms of plagiarism) was often deemed acceptable, until students were caught engaging in the behavior and disciplined with consequences. Thus teaching students about ethics took on great urgency, to positively influence not just the student's behavior in the careers that they would embark on, but more immediately during their time at the university.

Table 2: Computer engineering courses that integrate computing ethics modules at Colorado State University (2022-23)

| Course name | Level | Enrollment | Topic Coverage |
|---|---|---|---|
| ECE452: Computer Organization and Architecture | Undergraduate (Junior, Senior); Graduate | 70 | Inequality, justice, & human rights; semiconductor ethics; philosophy; codes of ethics |
| ECE554: (Advanced) Computer Architecture | Undergraduate (Senior); Graduate | 31 | Social & environmental impact; hardware & software security; philosophy; codes of ethics |
| CS/ECE561: Hardware/ Software Design of Embedded Systems | Undergraduate (Senior); Graduate | 51 | Privacy; design for safety; civic responsibility; research ethics; philosophy; codes of ethics |
| CS/ECE528: Embedded Systems and Machine Learning | Undergraduate (Senior); Graduate | 65 | AI & algorithms; bias; privacy and security; law and policy; future of work; philosophy |

Over the past year and a half (2022-2023), we have added short modules on computing ethics across core courses in our computer engineering curricula. Table 2 lists four courses into which these modules were integrated along with course level and student enrollment data. The courses are taken by junior and senior undergraduates, as well as graduate students in the departments of Electrical and Computer Engineering and Computer Science, at Colorado State University. Almost all courses include coverage of ethical theories and philosophy, as well as codes of ethics for the reasons discussed earlier. However, the topics that receive major emphasis vary across the courses. All courses collectively cover the main themes that were highlighted in Table 1 and its related discussion in the previous section. By integrating ethics across the computing curriculum in this manner, students are made aware of how ethical issues impact different areas in computing. The approach also exposes students to concrete ethical challenges in many real-world situations and provides them with repeated experiences of reasoning through ethical dilemmas.

Beyond the close integration of ethics with technical content across courses, our pedagogical approach relies heavily on and benefits from the case method [37]. Several real-world case studies are discussed, including the Pentium chip FDIV bug, the Therac-25 radiation disaster, the Toyota unintended acceleration bug, "Gaydar" AI, and the Ariane 5 rocket failure. These high profile cases are supplemented by the relatively more mundane, but more likely to be encountered real cases considered by the National Society of Professional Engineers (NSPE) Board of Ethical Review involving conflicts of interest, trade secrets, whistleblowing, and gift giving [38]. The coverage of these cases was intended to encourage students to express their ethical opinions in the classroom, formulate and justify decisions across practical contexts of ethics, and learn to effectively communicate on these themes.

More broadly, our intended learning outcomes from these ethics modules are manifold: provide the informational component to raise awareness of professional codes, ethical theories, and ethical challenges in computing; increase the ability to identify ethical issues associated with computing in real world scenarios; provide practice and training on ethical decision making, and inspire ethical behavior. The inspiration component of ethical pedagogy is often underappreciated [28]. One way students can be inspired to behave ethically is to tell them about professionals who made great sacrifices to do the right thing. Stories of individual courage displayed by exemplary engineers and scientists in the face of pressure to engage in wrongdoing can show students that it is possible to succeed in protecting the wellbeing of others, despite the personal risk involved. The students are likely to find themselves in tough situations in the computing profession, where they may feel intimidated (for possibly many reasons) to express their ethical concerns. They may even find themselves wanting to do the right thing but not have the moral courage to act in difficult situations. Inspiration can be a crucial cog in promoting that moral courage amongst these future computing professionals.

In course feedback questionnaires given to students for the ECE554 and CS/ECE528 courses taught in 2022, we deliberately decided not to include any specific questions asking students about the effectiveness of the ethics modules. From reports of such surveys in ethics courses in other universities, it has been shown that almost all students seem to indicate that they value ethics content. But this information does not allow us to determine how effective the ethics pedagogy actually was. Instead we were interested to see if students would talk about the ethics modules in the open ended comments section of the questionnaire without being prompted about it. In the past, students have expressed their opinions about the course, the instructor, and the content in this section of the survey. Interestingly, there were several comments about the ethics modules, all positive to varying degrees. One student in ECE554 said "I have [been] forced to partake in so many boring/unhelpful ethics lectures in the past, but the one[s] given in this course blew my mind. I highly recommend that one week to be open to the whole department to attend." In CS/ECE528, a student commented "I never realized ethical challenges with machine learning until you pointed so many of them in your lectures. More courses should talk about ethics." Another student in CS/ECE528 pointed out "I enjoyed the discussion and debate on ethics. The one change I would recommend is to more prominently highlight women engineers and scientists in the case studies. Almost all the case studies involved engineers who were men." As the case studies were taken from real cases in the past involving engineers and scientists, this comment highlighted how the computing profession has sadly been lacking in diversity. But it also emphasizes the opportunity, through including hypothetical case studies and performing more careful selection of cases that feature women and other traditionally under-represented groups in computing, to provide a more inclusive way to discuss ethics that could better inspire students of all backgrounds.

## 5. CONCLUSION

In this paper, we discussed the importance of teaching ethics in computer engineering and computer science curricula. We have highlighted how ethics is addressed in curricula across universities, and some of the key questions that impact the manner in which ethics should be taught. We then presented a case study of an ongoing effort that is integrating ethics modules across courses in the computing curricula at Colorado State University. Initial results and feedback have been very positive and motivate us to expand this approach to courses beyond the ones currently being considered, including electrical engineering courses. In the future, we plan to experiment with different pedagogical approaches to more effectively teach ethics modules, and improve inclusivity in the course content. We hope that our work can encourage instructors and developers of computing programs at universities to make ethics education more accessible to their students.